\documentclass[letter]{aa} 
\usepackage{graphicx}
\usepackage{txfonts}
\usepackage{longtable}
\usepackage{rotating}
\usepackage{graphics}
\usepackage{psfrag}
\usepackage{amssymb}
\def\bz{\left< B_{\rm z} \right>}
\def\nz{\left< N_{\rm z} \right>}

\begin{document}
   \title{
B fields in OB stars (BOB): FORS\,2 spectropolarimetric follow-up of the two rare rigidly rotating magnetosphere
stars HD\,23478 and HD\,345439\thanks{Based
on observations obtained in the framework of the ESO Prgs.\ 191.D-0255(E) and 094.D-0355(B).}}
\titlerunning{Fast rotating stars with magnetospheres}
\author{
S.~Hubrig\inst{1}
\and M.~Sch\"oller\inst{2}
\and L.~Fossati\inst{3}
\and T. Morel\inst{4}
\and N.~Castro\inst{3}
\and L.~M.~Oskinova\inst{5}
\and N.~Przybilla\inst{6}
\and S.~S.~Eikenberry\inst{7}
\and M.-F.~Nieva\inst{6}
\and N.~Langer\inst{3}
\and the BOB collaboration
}
\authorrunning{Hubrig et al.}

\institute{Leibniz-Institut f\"ur Astrophysik Potsdam (AIP), An der Sternwarte 16, 14482 Potsdam, Germany\\
              \email{shubrig@aip.de}
\and European Southern Observatory, Karl-Schwarzschild-Str.~2, 85748~Garching, Germany
\and Argelander-Institut f\"ur Astronomie, Universit\"at Bonn, Auf dem H\"ugel~71, 53121~Bonn, Germany
\and Institut d'Astrophysique et de G\'eophysique, Universit\'e de Li\`ege, All\'ee du~6~Ao\^ut, B\^at.~B5c, 
4000~Li\`ege, Belgium
\and Universit\"at Potsdam, Institut f\"ur Physik und Astronomie, 14476~Potsdam, Germany 
\and Institute for Astro- and Particle Physics, University of Innsbruck, Technikerstr.~25/8, 6020~Innsbruck, Austria
\and Department of Astronomy, University of Florida, 2011 Bryant Space Center, Gainesville, FL 32611
}
\date{Received; accepted}

  \abstract
   {}
   {
Massive B-type stars with strong
magnetic fields and fast rotation are very rare and provide a mystery for theories of both 
star formation and magnetic field evolution.
Only two such stars, called $\sigma$\,Ori\,E analogs, were previously known.
Recently, a team involved in APOGEE, one of the Sloan Digital Sky Survey~III programs, announced the discovery of two 
additional rigidly rotating magnetosphere stars, HD\,23478 and HD\,345439.
The presence of magnetic fields in these newly discovered $\sigma$\,Ori\,E analogs was not investigated in the past.  
}
   {
 In the framework of our ESO Large Programme, and one normal ESO programme, we carried out low-resolution FORS\,2
spectropolarimetric observations of HD\,23478 and HD\,345439.
}
   {
From the measurements using hydrogen lines, we discover a rather strong longitudinal magnetic field of the order of 
up to 1.5\,kG in HD\,23478, and up to 1.3\,kG using the entire spectrum.
The analysis of HD\,345439 using four subsequent spectropolarimetric subexposures does not reveal
the presence of a magnetic field at a significance level of 3$\sigma$.
On the other hand, the inspection of individual subexposures indicates that HD\,345439
may host a strong magnetic field, rapidly varying over 88\,minutes.
A hint at the fast rotation of HD\,345439 is also given by the behaviour of several metallic and \ion{He}{i} lines
in the low-resolution FORS 2 spectra, showing profile variations
already on such a short time scale.
}
   {}


   \maketitle

\section{Introduction}
The increasing rate of detections of magnetic fields in massive stars during the last years generates a 
strong motivation to study correlations between evolutionary state, rotation
velocity, and surface composition (e.g. Brott et al.\ \cite{Brott2011}; Potter et 
al.\ \cite{Potter2012}) aiming at understanding the origin and the role of magnetic fields 
in massive stars. 
Recent observations indicate that the presence of magnetic 
fields may be responsible for a wide range of phenomena observed in massive stars, such as chemical 
peculiarity, periodic UV wind-line variability, cyclic variability in H$\alpha$ and He~II 
$\lambda$4686, excess emission in UV-wind lines, and unusual X-ray emission 
(e.g.\ Schnerr et al.\ \cite{Schnerr2008}; Oskinova et al.\ \cite{Oskinova2011a}).
The variability is likely due to material that is guided by the large-scale magnetic field that
corotates with the star (e.g.\ Groote \& Hunger \cite{groote1982}).
To properly understand and model all these phenomena, the knowledge of the magnetic field behaviour,
i.e.\ its configuration, strength, and variability,
in massive stars is of utmost importance.

The presence of a rigidly rotating magnetosphere in the early B-type stars HD\,23478 and HD\,345439
was recently discovered in the Apache Point Observatory Galactic Evolution Experiment 
(APOGEE; Eikenberry et al.\ \cite{Eikenberry2014})
using high-resolution ($R\sim22\,500$) near-infrared H-band spectra. The authors detected in the 
APOGEE bandpass prominent Brackett series emission lines with broad double-horned profiles,
which are characteristic for the presence of a rigidly rotating
magnetosphere (RRM), like found in the prototype star
$\sigma$\,Ori\,E (see their Fig.~4).
The strong magnetic field and fast rotation lead to the formation of a centrifugally supported
magnetosphere with rotationally modulated hydrogen line emission explained within the RRM model
(e.g.\ Townsend et al.\ \cite{Townsend2005}).
Among $\sigma$\,Ori\,E, HD\,23478, and HD\,345439, the latter appears to display the strongest RRM feature. 
Eikenberry et al.\ (\cite{Eikenberry2014}) report that 
the optical spectra of HD\,23478 and HD\,345439 obtained with the Hobby-Eberly-Telescope at a medium resolving
power of 18\,000 indicate a classification B3IV for the former star and B1V--B2V for the latter.
While they determine a $v\,\sin\,i$ value for HD\,23478 of about $125\pm20$\,km\,s$^{-1}$,
HD\,345439, in contrast, appears to be one of the fastest known rotators among magnetic main-sequence stars with 
$v\,\sin\,i\sim270\pm20$\,km\,s$^{-1}$. Furthermore, the authors report that HD\,23478 exhibits a
broad H$\alpha$ RRM signature, while for HD\,345439 the H$\alpha$ profile shows a slight bump on the red 
side.

In this work, we present spectropolarimetric follow-up observations of
HD\,23478 and HD\,345439 obtained in 
the framework of our BOB (B fields in OB stars) collaboration (Morel et al.\ \cite{morel2014})
using FORS\,2 (FOcal Reducer low dispersion
Spectrograph) mounted on the 8-m Antu telescope of the VLT.  
We intend to search for the presence of strong magnetic fields in both stars, which
were implicitly detected through the RRM interpretation of their NIR spectra by 
Eikenberry et al.\ (\cite{Eikenberry2014}).

\section{Magnetic field measurements using FORS\,2 spectropolarimetry}

Low-resolution FORS\,2 observations for HD\,23478 and HD\,345439 were obtained in 2014 on November 17 and 
June 2, respectively. FORS\,2 is a multi-mode instrument equipped with polarisation analysing optics
comprising super-achromatic half-wave and quarter-wave phase retarder plates,
and a Wollaston prism with a beam divergence of 22$\arcsec$ in standard
resolution mode. 
We employed the GRISM 600B and the narrowest available slit width
of 0$\farcs$4 to obtain a spectral resolving power of $R\sim2000$.
The use of the mosaic detector 
with a  pixel size of 15\,$\mu$m allowed us to cover a
spectral range from 3250 to 6215\,\AA{}, which includes all Balmer lines 
apart from H$\alpha$, and numerous \ion{He}{i} lines.
For each star, four consecutive observations were obtained with pairs of position angles 
separated by 90$^{\circ}$, namely [$-45^{\circ}, +45^{\circ}, +45^{\circ}, -45^{\circ}, 
-45^{\circ}, +45^{\circ}, +45^{\circ}, -45^{\circ}$]. The exposure time for each position angle 
was 20\,sec for HD\,23478, and 600\,sec for HD\,345439.
A peak signal-to-noise ratio (SNR) of 1900 in the final 
Stokes $I$ spectrum was achieved for HD\,23478, while for HD\,345439 we obtained a peak SNR of $\sim$900.

The $V/I$ spectrum is calculated using:
\begin{equation}
\frac{V}{I} = \frac{1}{2} \left\{ 
\left( \frac{f^{\rm o} - f^{\rm e}}{f^{\rm o} + f^{\rm e}} \right)_{-45^{\circ}} -
\left( \frac{f^{\rm o} - f^{\rm e}}{f^{\rm o} + f^{\rm e}} \right)_{+45^{\circ}} \right\}
\end{equation}
where $+45^{\circ}$ and $-45^{\circ}$ indicate the position angle of the
retarder waveplate and $f^{\rm o}$ and $f^{\rm e}$ are the ordinary and
extraordinary beams, respectively.
Null profiles, $N$, are calculated as pairwise differences from all available 
$V$ profiles.  From these, 3$\sigma$-outliers are identified and used to clip 
the $V$ profiles.  This removes spurious signals, which mostly come from cosmic
rays, and also reduces the noise.
The mean longitudinal magnetic field, $\left< B_{\rm z}\right>$, is 
measured on the rectified and clipped spectra based on the relation
\begin{eqnarray} 
\frac{V}{I} = -\frac{g_{\rm eff}\, e \,\lambda^2}{4\pi\,m_{\rm e}\,c^2}\,
\frac{1}{I}\,\frac{{\rm d}I}{{\rm d}\lambda} \left<B_{\rm z}\right>\, ,
\label{eqn:vi}
\end{eqnarray} 

\noindent 
where $V$ is the Stokes parameter that measures the circular polarization, $I$
is the intensity in the unpolarized spectrum, $g_{\rm eff}$ is the effective
Land\'e factor, $e$ is the electron charge, $\lambda$ is the wavelength,
$m_{\rm e}$ is the electron mass, $c$ is the speed of light, 
${{\rm d}I/{\rm d}\lambda}$ is the wavelength derivative of Stokes~$I$, and 
$\left<B_{\rm z}\right>$ is the mean longitudinal (line-of-sight) magnetic field.
\begin{small}
\begin{table*}
\centering
\caption{
Summary of the results obtained for HD\,23478 and HD\,345439 from the analysis of the hydrogen 
lines and of the entire spectrum.
}
\label{tab:Bvalues}
\begin{tabular}{lcl|r@{$\pm$}rrr@{$\pm$}rr|r@{$\pm$}rrr@{$\pm$}rr}
\hline
\hline
\multicolumn{1}{c}{Object} &
\multicolumn{1}{c}{Subexposures} &
\multicolumn{1}{c|}{Group} &
\multicolumn{6}{c|}{Hydrogen} &
\multicolumn{6}{c}{All} \\
 &
 &
 &
\multicolumn{2}{c}{$\bz$ [G] } &
\multicolumn{1}{c}{$\sigma_B$} &
\multicolumn{2}{c}{$\nz$ [G] } &
\multicolumn{1}{c|}{$\sigma_N$} &
\multicolumn{2}{c}{$\bz$ [G] } &
\multicolumn{1}{c}{$\sigma_B$} &
\multicolumn{2}{c}{$\nz$ [G] } &
\multicolumn{1}{c}{$\sigma_N$} \\
\hline
HD\,23478  & all & Bonn    & $-$1477 &  95 & 15.5 &     30 &  64 & 0.5    & $-$1302 &  59 & 21.9 &    74 & 49 & 1.5  \\
           &     & Potsdam & $-$1347 & 114 & 11.8 &     50 &  93 & 0.5    & $-$1139 &  84 & 13.5 &    45 & 70 & 0.6  \\
\hline
HD\,345439 & all & Bonn    &       4 & 121 & 0.0  &     23 & 111 & 0.2    &    $-$2 &  78 &  0.0 &    36 & 74 & 0.5  \\
           &     & Potsdam &      57 & 153 & 0.4  &     58 & 136 & 0.4    &   $-$22 & 116 &  0.2 &    64 & 103 & 0.6  \\
\hline
     & 1+2 & Bonn    &     649 & 169 & 3.8  & $-$117 & 157 & 0.7    &     389 & 105 &  3.7 &   104 & 101 & 1.0 \\
           &     & Potsdam &     638 & 183 & 3.5  & $-$197 & 188 & 1.0    &     507 & 134 &  3.8 & $-$93 & 143 & 0.6 \\
\hline
     & 3+4 & Bonn    &  $-$618 & 171 & 3.6  &    188 & 154 & 1.2    &  $-$375 & 109 &  3.4 &    18 & 103 & 0.2 \\
           &     & Potsdam &  $-$574 & 193 & 3.0  &    281 & 192 & 1.5    &  $-$501 & 146 &  3.4 &   207 & 145 & 1.4 \\
\hline
           & 1   & Bonn    &     435 & 244 & 1.8  & \multicolumn{3}{c|}{} &     418 & 159 &  2.6 & \multicolumn{3}{c}{} \\
           &     & Potsdam &     414 & 282 & 1.5  & \multicolumn{3}{c|}{} &     436 & 212 &  2.1 & \multicolumn{3}{c}{} \\
\hline
     & 2   & Bonn    &     631 & 253 & 2.5  & \multicolumn{3}{c|}{} &     252 & 148 &  1.7 & \multicolumn{3}{c}{} \\
           &     & Potsdam &     789 & 246 & 3.2  & \multicolumn{3}{c|}{} &     565 & 188 &  3.0 & \multicolumn{3}{c}{} \\
\hline
     & 3   & Bonn    &  $-$509 & 242 & 2.1  & \multicolumn{3}{c|}{} &  $-$402 & 160 &  2.5 & \multicolumn{3}{c}{} \\
           &     & Potsdam &  $-$303 & 282 & 1.1  & \multicolumn{3}{c|}{} &  $-$298 & 212 &  1.4 & \multicolumn{3}{c}{} \\
\hline
           & 4   & Bonn    & $-$1181 & 328 & 3.6  & \multicolumn{3}{c|}{} &  $-$534 & 166 &  3.2 & \multicolumn{3}{c}{} \\
           &     & Potsdam &  $-$840 & 262 & 3.2  & \multicolumn{3}{c|}{} &  $-$689 & 198 &  3.5 & \multicolumn{3}{c}{} \\
\hline
\end{tabular}
\end{table*}
\end{small}
\begin{figure}
\centering
\includegraphics[width=0.45\textwidth]{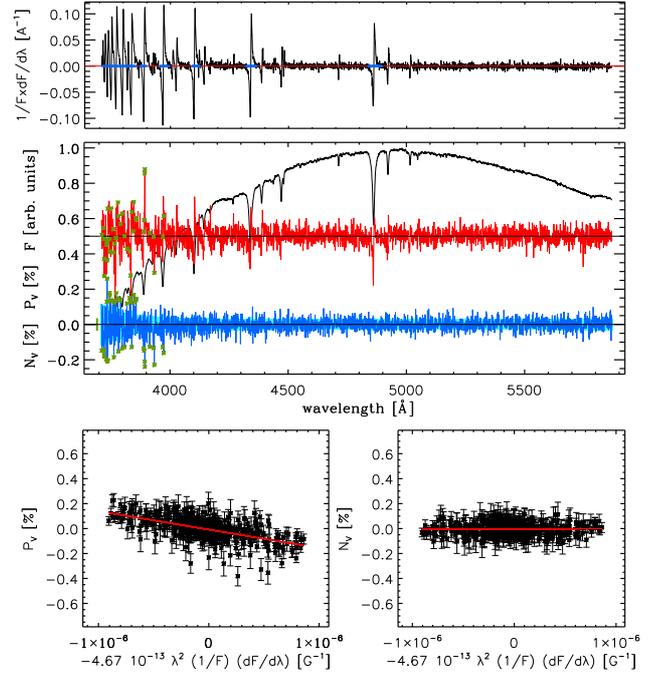}
\caption{Graphical output of the {\sc iraf}/{\sc idl} ``Bonn pipeline'' obtained from the analysis of 
the FORS\,2 data of HD\,23478 considering the hydrogen lines. The top panel presents the derivative of Stokes~$I$. 
The regions used for the calculation of the magnetic field are marked by a thick blue line centered at zero. 
In the middle panel, the top spectrum presents the Stokes~$I$ spectrum 
arbitrarily normalised to the highest value, the middle red 
spectrum corresponds to the Stokes~$V$ spectrum (in \%), while the bottom blue spectrum is the $N$ spectrum (in \%). The 
green asterisks drawn on the Stokes~$V$ and $N$ spectra mark the points which have been removed by the
sigma clipping. The pale blue strip drawn behind the $N$ spectrum shows the uncertainty 
associated with each spectral point. As a double-check of the statistical consistency of the 
data and uncertainties, the thick green bar on the left side of the $N$ spectrum shows the standard 
deviation of the $N$ profile. In the bottom-left panel, we present the linear fit used for the determination of the magnetic 
field value using Stokes~$V$ (i.e.\ $\bz$). The red solid line shows the best fit. From the linear fit we 
obtain $\bz=-1477\pm95$\,G. In the bottom-right panel, we present the same as in the bottom-left panel, but 
for the null profile (i.e.\ $\nz$). From the linear fit, we obtain $\nz=30\pm64$\,G.}
\label{fig:HD23478}
\end{figure}

The longitudinal magnetic field is usually measured in two ways: using only the hydrogen Balmer
lines or using the entire spectrum including all available lines.
To identify any systematic differences that could exist in treating the FORS\,2 data by different research groups,
the mean longitudinal 
magnetic field, $\bz$, was derived in both stars using independent software packages (one developed in Bonn and the other one in Potsdam). 
For the first reduction, we used a suite of IRAF 
(Tody \cite{Tody1993})\footnote{Image Reduction and Analysis 
Facility (IRAF -- {http://iraf.noao.edu/}) is distributed by the National Optical Astronomy 
Observatory, which is operated by the Association of Universities for Research in Astronomy (AURA) 
under cooperative agreement with the National Science Foundation.} 
and IDL routines that follow the technique, recipes, and recommendations by Bagnulo et al.
(\cite{bagnulo2002,bagnulo2012,bagnulo2013})\footnote{More details about the applied data reduction and analysis
 procedure of FORS spectropolarimetric data will be given in a separate work (Fossati et al., in preparation).}.
The determination of the mean longitudinal magnetic field using low-resolution FORS spectropolarimetry with the 
second software package developed in Potsdam is described by
Hubrig et al.\ (\cite{Hubrig2014,Hubrig2015}) and by Sch\"oller et al.\ (in preparation). 
In general, the measurement method is the same for both pipelines. 
A few minor differences refer 
to the clipping procedure, rectification, and the choice of the wavelength
regions including the hydrogen lines.
Furthermore, Monte Carlo bootstrapping tests are carried out in the second software package (e.g. 
Rivinius et al.\ \cite{rivinius2010}).
In these tests, we generate 250\,000 statistical variations of the 
original dataset by the bootstrapping technique, and analyze the resulting 
distribution $P(\left<B_{\rm z}\right>)$ of the regression results.
Mean and standard deviation of this distribution are identified with the most 
likely mean longitudinal magnetic field and its $1\,\sigma$ error, 
respectively. 
The main advantage of this method is that it provides an 
independent error estimate.
The measurement uncertainties for both stars obtained before and after Monte Carlo bootstrapping tests were found to be 
in close agreement, indicating the robustness of the measurement method. 

The results of the magnetic field measurements carried out using independent reduction and measurement packages
are listed in Table~\ref{tab:Bvalues}. For the first software package developed in Bonn, the uncertainty 
on the $\bz$ and $\nz$ values is the $\chi^2$-scaled uncertainty (Bagnulo et al.\ \cite{bagnulo2012}), 
while the uncertainty determined by the second software package developed in Potsdam involves in addition the bootstrapping results. 
In each column, the numbers $\sigma_B$ and $\sigma_N$ after the $\bz$ and $\nz$ values refer to the sigma detection. 

A strong magnetic field at a significance level of more than 10$\sigma$ was achieved for HD\,23478 
using the hydrogen lines and the entire spectrum. Using the first software package we obtain 
$\bz=-1302\pm59$\,G for the whole spectrum and 
$\bz_{\rm hyd}=-1477\pm95$\,G,
while using the second software package we measure $\bz=-1139\pm84$\,G  and 
$\bz_{\rm hyd}=-1347\pm114$\,G. For both reduction packages the magnetic field values obtained for 
each pair of position angles
separated by 90$^{\circ}$ show negative polarity, are all consistent with each other, and are of similar order.
No significant fields were detected in the null spectra 
calculated by combining the subexposures in such a way that the intrinsic source polarization cancels out,
yielding a diagnostic null $N$ spectrum.
Figure~\ref{fig:HD23478} presents the main graphical output of the ``Bonn pipeline'' showing the 
clear magnetic field detection obtained form the analysis of the hydrogen lines of HD\,23478.

\begin{figure}
\centering
\includegraphics[width=0.45\textwidth]{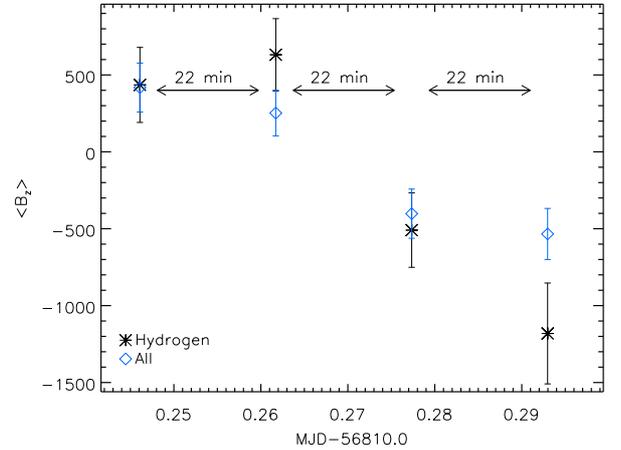}
\caption{Variation of the $\bz$ value obtained for HD\,345439 from the analysis of the hydrogen 
lines (black asterisks) and of the whole spectrum (blue rhombs). Between each point there is a 
time span of about 22\,min.}
\label{fig:HD345439c}
\end{figure}

\begin{figure*}
\centering
\includegraphics[width=0.40\textwidth]{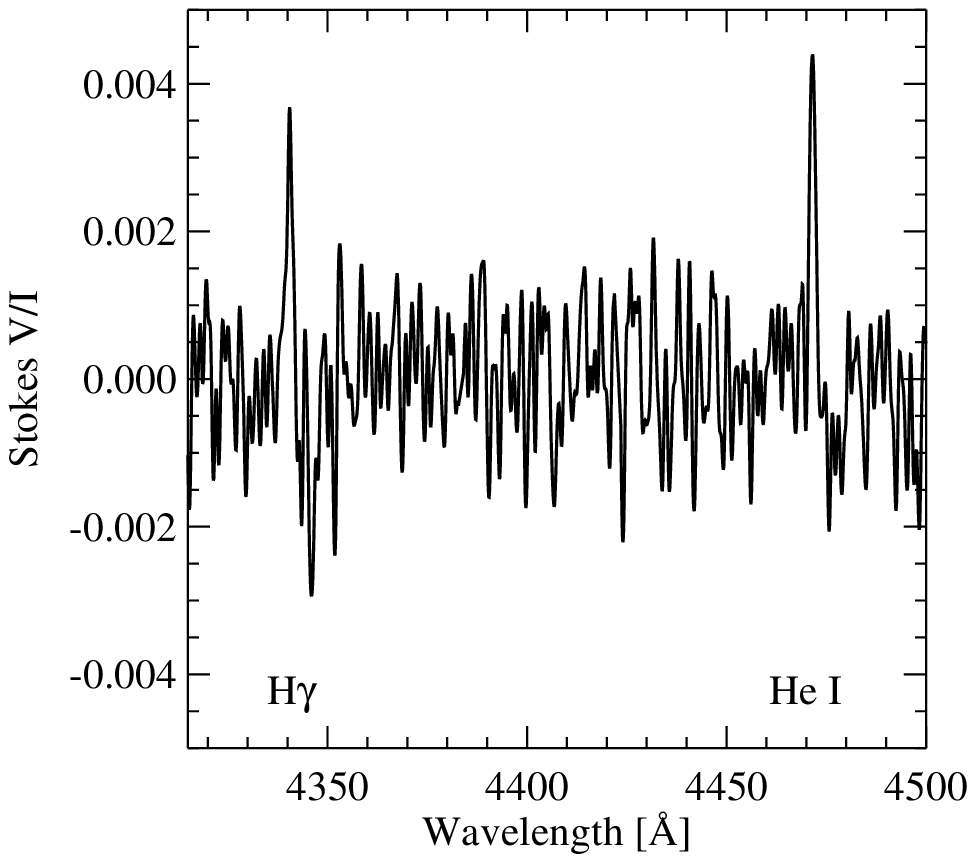}
\includegraphics[width=0.40\textwidth]{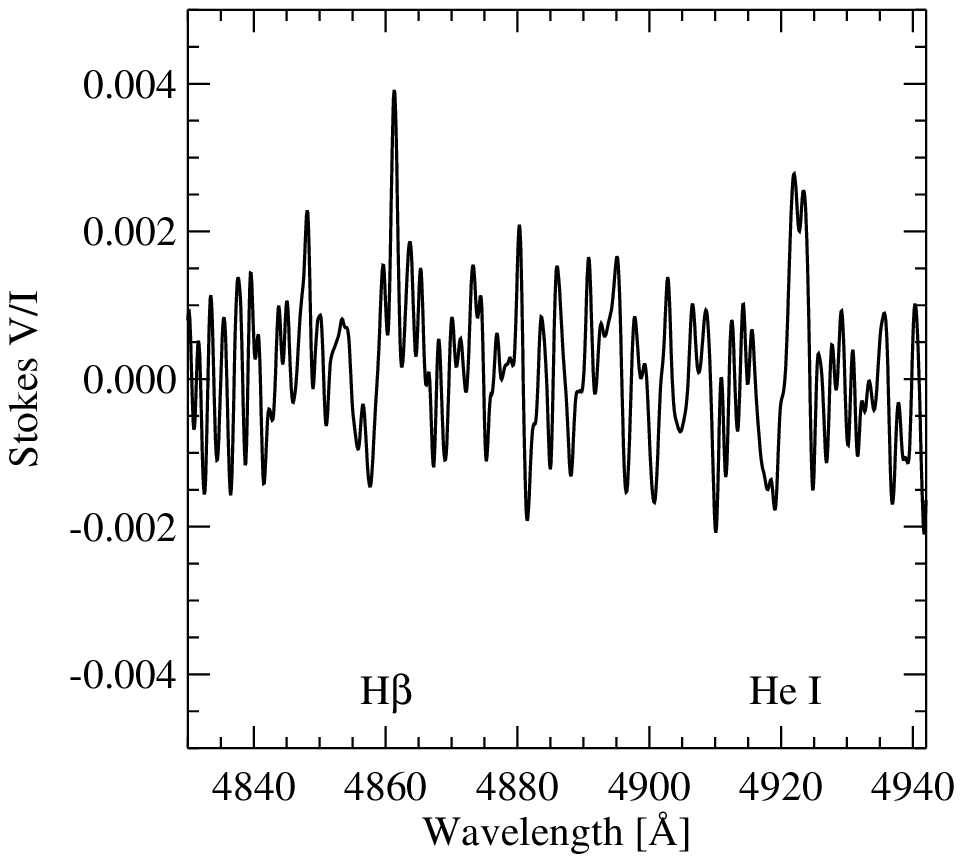}
\caption{
Left-side panel: Stokes~$V$ spectrum of HD\,345439 in the vicinity of H$\gamma$ and \ion{He}{i}  
$\lambda$4471 obtained from the combination of the first two pairs of subexposures. The shape of the
Zeeman features indicates the presence of a longitudinal magnetic field of positive polarity.
Right-side panel: Stokes~$V$ spectrum of HD\,345439 in the vicinity of H$\beta$ and \ion{He}{i}  
$\lambda$4922 lines obtained from the combination of the last two pairs of subexposures. 
The shape of the Zeeman features indicates the presence of a longitudinal magnetic field of negative polarity.
}
\label{fig:HD345439mf}
\end{figure*}

In the analysis of HD\,345439, using four consecutive observations obtained with pairs of position angles 
separated by 90$^{\circ}$,
we obtained a clear non-detection, which is presented in the second 
line in Table~\ref{tab:Bvalues}. However, after splitting the observations into two data sets,
i.e.\ using the first two pairs and the second two pairs consisting 
of observations at the retarder waveplate positions [$-45^{\circ}, +45^{\circ}, +45^{\circ}, -45^{\circ}$],
we obtain 3.0 to 3.8$\sigma$ detections 
shown in the third and fourth lines of the table. The fact that the first and second 
data-sets led to opposite $\bz$ values, though almost equal in modulus, led to a non-detection in the measurements
using the whole available data set. 
This behaviour might be suggestive of a very fast rotating star 
with a magnetic field axis inclined relative to the rotation axis.
Given the apparent fast and large variation of the value of the longitudinal magnetic field, 
we then decided to measure the magnetic field using each individual pair obtained at the retarder waveplate positions
[$-45^{\circ}, +45^{\circ}$] and separated by roughly 22\,minutes. The results are presented in the last four 
lines of Table~\ref{tab:Bvalues}. In this case, the null profile can not be extracted, since 
the measurements were done using observations at just two position angles. Note also that no 
sigma clipping was adopted for this specific analysis, 
because the sigma clipping is applied on the basis of the null profile. Figure~\ref{fig:HD345439c} 
shows the variation of the $\bz$ value obtained from the analysis of the hydrogen lines and of the entire spectrum
using the first software package developed in Bonn. 
We discover that the mean longitudinal magnetic field measured on hydrogen lines is changing from 
about $+$500\,G measured in the first
pair of subexposures to about $-$1200\,G measured in the last pair of subexposures. Using the second 
software package developed in Potsdam we obtained
similar magnetic field behaviour, but with somewhat different absolute values of the field.
The detections are achieved at a significance level of 3 to 3.6$\sigma$ in two subexposures.
In Figure~\ref{fig:HD345439mf}, we present Stokes $V$ spectra obtained for the combinations
of the first two pairs of subexposures and for the last two pairs of subexposures in the vicinity of the hydrogen and
\ion{He}{i} lines. The different shape of the Zeeman features indicates that the magnetic field polarity has changed after the
the second subexposure.
We note that all tests we performed on the data did not give an indication 
of the presence of relevant systematic noise. Nevertheless, 
it is important to obtain in the future more 
spectropolarimetric data to confirm this behaviour.

\begin{figure}
\centering
\includegraphics[width=0.50\textwidth]{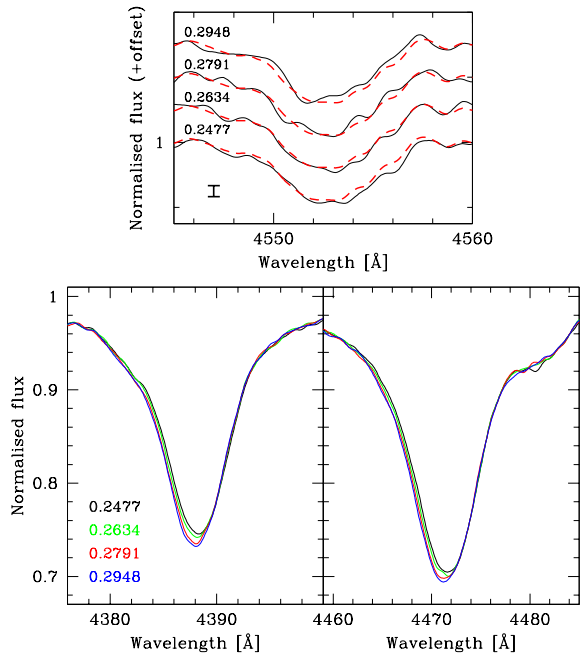}
\caption{Line profile variability of Si and He in the FORS\,2 Stokes $I$ spectra of HD\,345439
obtained for each pair of subexposures, i.e.\ with a time lapse of 22\,min.
Top panel presents the behaviour of the \ion{Si}{iii} $\lambda$4553 line. The mean profile is indicated by 
the red dashed line. The numbers close to the plotted line profiles refer to the time of the observation 
calculated as MJD-56810. The vertical bar on the left side shows the noise level.
In the bottom panel we present overplotted profiles of the \ion{He}{i} $\lambda$4388 and \ion{He}{i} $\lambda$4471 lines. 
Their intensity is lower
at the phase when the polarity of the field is positive and it becomes higher when the field is negative.
The error bar is too small to be presented in these plots.
}
\label{fig:var}
\end{figure}

An indication for the fast rotation of HD\,345439 is also given by the behaviour of 
several metallic and \ion{He}{I} lines in the low-resolution FORS\,2 spectra, showing 
profile variations already on a time scale of 88\,min.
In Fig.~\ref{fig:var}, we present the behaviour of the \ion{Si}{iii} $\lambda$4553 and 
the \ion{He}{i} $\lambda$4388 and \ion{He}{i} $\lambda$4471 lines in the 
FORS\,2 Stokes $I$ spectra obtained for each pair of subexposures recorded with a time lapse
of 22\,min. The profile of the \ion{Si}{iii} line becomes asymmetric in the last pair of subexposures,
i.e. at the time when we observe negative field polarity. We note, however, that the level of variability of 
the \ion{Si}{iii} $\lambda$4553 line profile is not significantly higher than that
in neighbouring continuum regions. On the other hand, similar asymmetric profiles of Si lines are detected in our 
FEROS (Fiber-fed Extended Range Optical Spectrograph) spectrum of $\sigma$\,Ori\,E obtained in the framework 
of the ESO programme 091.C-0713(A).
As for the behaviour of the \ion{He}{i} $\lambda$4388 and \ion{He}{i} $\lambda$4471 lines,
their intensity is lower
at the phase when the polarity of the field is positive and it becomes higher when the field is negative. 
Such a fast spectral and longitudinal magnetic field variability suggests that the rotation period should be
very short, probably less than one day. The rapid change in the strength of the mean longitudinal magnetic 
field by about 1\,kG over 88\,min in HD\,345439 is reminiscent of the mean longitudinal magnetic field behaviour 
in the fast rotating 
$\sigma$\,Ori\,E analog HR\,7355 with a rotation period of 0.52\,d. This star shows a field variation of 
about 1\,kG
within a phase interval of 0.1, corresponding to 75\,min (see Fig.~3 in Rivinius et al.\ \cite{Rivinius2013}).
Therefore, it is not impossible that the rotation period of HD\,345439 is also significantly shorter than one day.

No spectral variability was detected in the eight low-resolution FORS\,2 spectra 
for HD\,23478 taken within 15\,min.
This star was 
intensively studied photometrically by Jerzykiewicz (\cite{Jerzykiewicz1993}), who suggested a photometric/rotation
period of 1.05\,d and, using a radius estimated from the spectral type, obtained 240\,km\,s$^{-1}$ for the
equatorial rotation velocity.

\section{Discussion}
\label{sect:disc}

Our spectropolarimetric follow-up of the early B-type stars HD\,23478 and HD\,345439, 
with recently discovered rigidly rotating magnetospheres, using FORS\,2 observations,
reveals the presence of a strong longitudinal magnetic field of the order of
1.5\,kG in HD\,23478.
We also found an indication for the presence of a rapidly changing magnetic field in the fast rotator HD\,345439.
Obviously, additional follow-up spectropolarimetric monitoring of the behaviour of the magnetic field in HD\,23478 and 
HD\,345439 is urgently needed to confirm and characterize the magnetic field topology of both stars.
Knowledge of the structure of the magnetic field is of crucial importance for the understanding of the
nature of these stars belonging to the group of $\sigma$\,Ori\,E analogs.

This type of stars is extremely rare:
apart from $\sigma$\,Ori\,E (e.g.\ Groote \& Hunger \cite{GrooteHunger1997}),
only two other stars, HR\,5907 and HR\,7355, are known to belong to 
this group called the $\sigma$\,Ori\,E analogs
(Grunhut et al.\ \cite{Grunhut2012}; Rivinius et al.\ \cite{Rivinius2013}).
Their very strong
magnetic fields and extraordinary fast rotation provide a mystery for theories of both 
star formation and magnetic field evolution.
Both stars have similar dipole strengths of the order of 10\,kG and similar very short half-day rotation periods. 
However, the spindown 
timescale via magnetic braking is expected to be much shorter than their estimated ages (e.g., 
Rivinius et al.\ \cite{Rivinius2013}; Mikul{\'a}{\v s}ek et al.\ \cite{Mikulasek2010}).

The emission variability in hydrogen lines in the $\sigma$\,Ori\,E analogs suggests that gas is trapped 
in magnetospheric clouds. The high strength of the
magnetic fields controlling the stellar winds and fast rotation 
makes these type of stars a test case for models of X-ray production. 
Only low resolution X-ray 
spectra were so far measured for three members of this group, $\sigma$\,Ori\,E, HR\,5907, and HR\,7355 
(e.g., Petit et al.\ \cite{Petit2013}; Naz\'e et al.\ \cite{Naze2014}). 
We retrieved and 
analyzed these archival {\em XMM-Newton} and {\em Chandra} observations using most recent calibrations. 
Our analysis of the X-ray spectra of HR\,5907 and HR\,7355 shows that these stars are even harder X-ray sources 
than $\sigma$\,Ori\,E; e.g., the emission measure weighted temperature of the hot plasma in HR\,5907 is 80\,MK,
and in HR\,7355 it is 60\,MK. Given that the winds in magnetic B-type dwarf stars are weak, with velocities not exceeding 
a few hundred km\,s$^{-1}$ (Oskinova et al.\ \cite{Oskinova2011b}), it is very difficult to explain such high temperatures 
in the framework of the magnetically confined wind shock model (Babel \& Montmerle \cite{BabelMontmerle1997}). On the other hand, 
non-thermal emission can be expected from stars with strong magnetic fields (Leto et al.\ \cite{Leto2006}). Our analysis 
of archival data showed that the presence of non-thermal radiation in these objects is very likely (Oskinova et al., in preparation).
New, already approved {\em XMM-Newton} observations of HD\,345439 will probe whether hard non-thermal X-rays 
are present in this object and shed new light on the mechanisms of X-ray generation in strongly magnetic 
fast rotators.

\begin{acknowledgements}
TM acknowledges financial support from Belspo for contract PRODEX GAIA-DPAC.
LF acknowledges financial support from the Alexander von Humboldt Foundation.
LMO thanks DLR grant 50\,OR\,1302. The authors thank to Rainer Arl, Artemio Herrero, and Fabian Schneider for 
useful comments.
\end{acknowledgements}


\begin{thebibliography}{}

\bibitem[1997]{BabelMontmerle1997}
Babel, J. \& Montmerle, T.\ 1997,
A\&A, 323, 121

\bibitem[2002]{bagnulo2002}
Bagnulo, S., Szeifert, T., Wade, G.~A., et al.\ 2002,
\aap, 389, 191

\bibitem[2012]{bagnulo2012}
Bagnulo, S., Landstreet, J.~D., Fossati, L. \& Kochukhov, O.\ 2012, 
\aap, 538, A129

\bibitem[2013]{bagnulo2013}
Bagnulo, S., Fossati, L., Kochukhov, O. \& Landstreet, J.~D.\ 2013, 
\aap, 559, A103

\bibitem[2011]{Brott2011}
Brott, I., Evans, C.~J., Hunter, I., et al.\ 2011,
A\&A, 530, A116 

\bibitem[2014]{Eikenberry2014}
Eikenberry, S.~S., Chojnowski, S.~D., Wisniewski, J., et al.\ 2014,
ApJL, 784, L30

\bibitem[1982]{groote1982}
Groote, D., \& Hunger, K.\ 1982,
A\&A, 116, 64

\bibitem[1997]{GrooteHunger1997}
Groote, D., \& Hunger, K.\ 1997,
\aap, 319, 250

\bibitem[2012]{Grunhut2012}
Grunhut, J.~H., Rivinius, Th., Wade, G.~A., et al.\ 2012,
MNRAS, 419, 1610

\bibitem[2014]{Hubrig2014}
Hubrig, S., Sch\"oller, M., \& Kholtygin, A.~F.\ 2014,
MNRAS, 440, 1779

\bibitem[2015]{Hubrig2015}
Hubrig, S., Sch\"oller, M., Kholtygin, A.~F., et al.\ 2015,
MNRAS, 447, 1885

\bibitem[1993]{Jerzykiewicz1993}	
Jerzykiewicz, M.\ 1993,
A\&AS, 97, 421

\bibitem[2006]{Leto2006}
Leto, P., Trigilio, C., Buemi, C.~S., et al.\ 2006,
A\&A, 458, 831

\bibitem[2010]{Mikulasek2010}
Mikul{\'a}{\v s}ek, Z., Krti{\v c}ka, J., Henry, G.~W., et al.\ 2010,
A\&A, 511, L7

\bibitem[2014]{morel2014}
Morel, T., Castro, N., Fossati, L., et al.\ 2014,
Msngr. 157, 27

\bibitem[2014]{Naze2014}
Naz\'e, Y., Petit, V., Rinbrand, M., et al.\ 2014,
ApJS, 215, 10

\bibitem[2011a]{Oskinova2011a}
Oskinova, L.~M., Hamann, W.-R., Cassinelli, J.~P., et al.\ 2011a,
AN, 332, 988

\bibitem[2011b]{Oskinova2011b}
Oskinova, L.~M., Todt, H., Ignace, R., et al.\ 2011b,
MNRAS, 416, 1456

\bibitem[2013]{Petit2013}
Petit, V., Owocki, S.~P., Wade, G.~A., et al.\ 2013,
MNRAS, 429, 398

\bibitem[2012]{Potter2012}
Potter, A.~T., Chitre, S.~M., Tout, C.~A.\ 2012,
MNRAS, 424, 2358

\bibitem[2010]{rivinius2010}
Rivinius, T., Szeifert, T., Barrera, L., et al. 2010, MNRAS, 405, L46 

\bibitem[2013]{Rivinius2013}
Rivinius, T., Townsend, R.~H.~D., Kochukhov, O., et al.\ 2013,
MNRAS, 429, 177

\bibitem[2008]{Schnerr2008}
Schnerr, R.~S., Henrichs, H.~F., Neiner, C., et al.\ 2008,
A\&A, 483, 857 

\bibitem[1993]{Tody1993}
Tody, D.\ 1993,
in Astronomical Data Analysis Software and Systems II,                   
ed.\ R.~J.~Hanisch, R.~J.~V. Brissenden, \& J.~Barnes (San Francisco: ASP),
ASP Conf.\ Ser., 52, 173

\bibitem[2005]{Townsend2005}
Townsend, R.~H.~D., Owocki, S.~P., \& Groote, D.\ 2005,
{\apjl}, 630, L81

\end{thebibliography}
\end{document}